\documentclass[twocolumn,showpacs,preprintnumbers,showkeys,superscriptaddress]{revtex4}
\usepackage{amssymb}
\usepackage{graphicx}
\usepackage{dcolumn}
\usepackage{bm}
\usepackage{appendix}

\def\eqref#1{Eq.~(\ref{eq:#1})}

\begin{document}

\title{Random Phase Approximation without Bogoliubov Quasi-particles}
\author{L. Y. Jia}   \email{jialiyuan84@gmail.com}
\affiliation{Department of Physics, University of Shanghai for
Science and Technology, Shanghai 200093, P. R. China}
\affiliation{Department of Physics, Hebei Normal University,
Shijiazhuang, Hebei 050024, P. R. China}

\date{\today}

\begin{abstract}

A new version of random phase approximation is proposed for
low-energy harmonic vibrations in nuclei. The theory is not based on
the quasi-particle vacuum of the BCS/HFB ground state, but on the
pair condensate determined in Ref. \cite{Jia_pairing}. The current
treatment conserves the exact particle number all the time. As a
first test the theory is considered in two special cases: the
degenerate model (large pairing limit) and the vanished-pairing
limit.

\end{abstract}


\vspace{0.4in}

\maketitle

\section{Introduction}

The random phase approximation (RPA) is first proposed in plasma
physics to describe plasma oscillations \cite{RPA_plasma}. In
low-energy nuclear structure usually the extended version,
quasi-particle random phase approximation (QRPA), is used owing to
the existence of pairing correlations \cite{Bohr, Peter}. In the
mean-field order, we introduce Bogoliubov quasi-particles for
pairing and solve the Hartree-Fock-Bogoliubov (HFB) equation
(advanced version of BCS), and the unperturbed ground state is
written as the quasi-particle vacuum. On top of that,
small-amplitude vibration is described by QRPA in terms of linear
combinations of two quasi-particle excitations.

The QRPA has problems inherent from its starting point -- BCS mean
field. The wavefunction does not have definite particle number,
which is conserved only on average. Also, for nuclei near the
critical pairing strength of BCS, the mean-field description fails
itself and certainly QRPA is inapplicable. Complicated projection
techniques to good particle number are needed to restore the broken
symmetry.

Recently we proposed a number-conserving theory for nuclear pairing
\cite{Jia_pairing}. Without introducing quasi-particles, the
unperturbed ground state is written as a $N$-pair condensate for a
system with $2N$ particles. Within a similar computing time to that
of BCS, we solve the pair structure, occupation numbers, and
pair-transition amplitudes. The theory is shown to be valid at
arbitrary pairing strength including those below the critical point
of BCS.

In this work we try to develop a new version of number-conserving
random phase approximation (N-RPA) based on the above
number-conserving ground state. The derivation of N-RPA is quite
straightforward following that of conventional (Q)RPA. In the end a
linear homogenous equation is resulted, and the frequency $\omega$
is determined by requiring the determinant to be zero. The computing
time cost is similar to that of conventional (Q)RPA. As a first
test, the theory is considered in two special cases: the degenerate
model (large pairing case) and the vanished-pairing case. In the
latter the theory goes over to the conventional RPA. Further tests
and applications are needed in the future.

\section{FORMALISM}

First we briefly repeat the essential result from Ref.
\cite{Jia_pairing} as the starting point of this work. The
antisymmetrized fermionic Hamiltonian is
\begin{eqnarray}
H = \sum_{12} \epsilon_{12} a_1^\dagger a_2 + \frac{1}{4}
\sum_{1234} V_{1234} a_1^\dagger a_2^\dagger a_3 a_4 .  \label{H_f}
\end{eqnarray}
The unperturbed ground state of the $2N$-particle system is assumed
to be a $N$-pair condensate,
\begin{eqnarray}
|\phi_N\rangle = \frac{1}{\sqrt{\chi_{N}}} (P^\dagger)^{N} |0\rangle
, \label{gs}
\end{eqnarray}
where $\chi_{N}$ is the normalization factor, and $P^\dagger$ is the
pair creation operator
\begin{eqnarray}
P^\dagger = \frac{1}{2} \sum_1 v_1 a_1^\dagger a_{\tilde{1}}^\dagger
. \label{P_dag}
\end{eqnarray}
By the main equation (19) in Ref. \cite{Jia_pairing}, we determine
the pair structure $v_1$, the density matrices
\begin{eqnarray}
\rho_{12} \equiv \langle \phi_{N} | a_2^\dagger a_1 | \phi_{N}
\rangle =
\delta_{12} n_1 ,  \label{rho_diag} \\
\kappa_{12} \equiv \langle \phi_{N-1} | a_2 a_1 | \phi_{N} \rangle =
\delta_{\tilde{1}2} s_1 , \label{kappa_diag}
\end{eqnarray}
and the mean fields
\begin{eqnarray}
f_{12} \equiv \epsilon_{12} + w_{12} \equiv \epsilon_{12} + \sum_{34} V_{1432} \rho_{34} = \delta_{12} e_1 ,  \label{f_diag} \\
\delta_{12} \equiv \frac{1}{2} \sum_{34} V_{1234} \kappa_{43} =
\delta_{1\tilde{2}} g_1 .     \label{delta_diag}
\end{eqnarray}

Now we begin the derivation of N-RPA. Here we do not include angular
momentum because in some cases (for example the deformed QRPA)
rotational symmetry is not conserved. We assume that the correlated
ground state $|0_N\rangle$ and the excited state $|1_N\rangle$ are
related by
\begin{eqnarray}
|1_N\rangle = A^\dagger |0_N\rangle ~,~~ A^\dagger = {\rm Tr}\{c R\}
= \sum_{12} c_{12} a_1^\dagger a_2 ,     \label{phonon}
\end{eqnarray}
where $R_{12} \equiv a_2^\dagger a_1$ are density matrix operators.
$c_{12}$ in the phonon creation operator $A^\dagger$ are parameters
to be determined later by the main equation (\ref{N_RPA}). The
correlated ground state $|0_N\rangle$ is not the unperturbed pair
condensate $|\phi_N\rangle$ (\ref{gs}), but has ``2-particle-2-hole
components'' in it.

Let us calculate the following quantity
\begin{eqnarray}
\langle 0_N | R_{12} | 1_N \rangle = \sum_{3} c_{13} \langle 0_N |
a_2^\dagger a_3 | 0_N \rangle  \nonumber \\
- \sum_{34} c_{43} \langle 0_N | a_2^\dagger a_4^\dagger a_1 a_3 |
0_N \rangle , \label{R_01_exact}
\end{eqnarray}
where we have used Eq. (\ref{phonon}). Here we make the main
approximation of the theory, the so-called ``linearization''. We
assume that
\begin{eqnarray}
\langle 0_N | a_2^\dagger a_1 | 0_N \rangle \approx \langle \phi_N |
a_2^\dagger a_1 | \phi_N \rangle = \rho_{12} ,     \label{fac_1b}
\end{eqnarray}
and
\begin{eqnarray}
\langle 0_N | a_4^\dagger a_3^\dagger a_2 a_1 | 0_N \rangle \approx
\langle \phi_N | a_4^\dagger a_3^\dagger a_2 a_1 | \phi_N \rangle
\nonumber \\
= \langle a_4^\dagger a_1 \rangle \langle a_3^\dagger
a_2 \rangle - \langle a_4^\dagger a_2 \rangle \langle a_3^\dagger
a_1 \rangle + \langle a_4^\dagger a_3^\dagger \rangle \langle a_2
a_1 \rangle
\nonumber \\
= \rho_{14} \rho_{23} - \rho_{24} \rho_{13} + (\kappa^\dagger)_{34}
\kappa_{12} ,     \label{fac_2b}
\end{eqnarray}
where $\langle a_3^\dagger a_2 \rangle$ means $\langle \phi_N |
a_3^\dagger a_2 | \phi_N \rangle$, $\langle a_2 a_1 \rangle$ means
$\langle \phi_{N-1} | a_2 a_1 | \phi_N \rangle$, etc. In Eqs.
(\ref{fac_1b}) and (\ref{fac_2b}) we made the usual approximation
(see Ref. \cite{Peter}) that the correlated ground state
$|0_N\rangle$ does not differ much from the unperturbed ground state
$|\phi_N\rangle$ if the vibrational amplitudes are small. Under the
approximations (\ref{fac_1b}) and (\ref{fac_2b}), Eq.
(\ref{R_01_exact}) becomes
\begin{eqnarray}
\langle 0_N | R_{12} | 1_N \rangle = c_{12} n_2 (1 - n_1) +
c_{\tilde{2}\tilde{1}} s_1 s_2 .     \label{R_01}
\end{eqnarray}

The exact Heisenberg equation of motion for the density matrix
operators $R_{12} = a_2^\dagger a_1$ is calculated as
\begin{eqnarray}
[R_{12}, H] = [\epsilon , R]_{12}  \nonumber \\
- \frac{1}{2} \sum_{345} V_{5432} a_5^\dagger a_4^\dagger a_3 a_1 +
\frac{1}{2} \sum_{345} V_{1345} a_2^\dagger a_3^\dagger a_4 a_5 .
\label{exact_eom}
\end{eqnarray}
We bra-ket Eq. (\ref{exact_eom}) with ``$\langle 0_N|$'' and ``$|1_N
\rangle$''. On the left-hand side we have $\langle 0_N | [R_{12}, H]
| 1_N \rangle = (E^1_N - E^0_N) \langle 0_N | R_{12} | 1_N \rangle$,
where $E^1_N - E^0_N \equiv \omega$ is the N-RPA frequency, and
$\langle 0_N | R_{12} | 1_N \rangle$ is already given in Eq.
(\ref{R_01}). The right-hand side is handled in a way similar to
that of Eqs. (\ref{R_01_exact})-(\ref{R_01}): after substituting Eq.
(\ref{phonon}), we again use Eqs. (\ref{fac_1b}), (\ref{fac_2b}),
and their extension to three-body density matrix operators $\langle
0_N | a_6^\dagger a_5^\dagger a_4^\dagger a_3 a_2 a_1 | 0_N
\rangle$, which is approximated by all possible ``fully contracted
terms'' (there are $15$ terms) in Wick's theorem (although normal
ordering of operators may be hard to define). In the end we get
\begin{widetext}
\begin{eqnarray}
\omega ~ [ c_{12} n_2 (1 - n_1) + c_{\tilde{2}\tilde{1}} s_1 s_2 ] =
\sum_3 \epsilon_{13} [c_{32} n_2 (1 - n_3) + c_{\tilde{2}\tilde{3}}
s_2 s_3] - \sum_3 [c_{13} n_3 (1 - n_1) + c_{\tilde{3}\tilde{1}} s_1
s_3] \epsilon_{32}     \nonumber \\
+ \sum_3 w_{13} [ c_{32} n_2 (1 - n_3) - c_{\tilde{2}\tilde{3}} s_2
s_3 ] - \sum_3 [ c_{13} n_3 (1 - n_1) - c_{\tilde{3}\tilde{1}} s_{1}
s_3 ] w_{32}
  \nonumber \\
- (n_{1} - n_{2}) \sum_{34} V_{1432} [ c_{34} n_{4} (1 - n_{3}) +
c_{\tilde{4}\tilde{3}} s_{3} s_4 ] + c_{12} [ g_2 s_2 (1 - n_1) +
g_1 s_1 n_2 ] + c_{\tilde{2}\tilde{1}} [ g_2 s_1(1-n_2) + g_1 s_2
 n_1 ]
\nonumber \\
+ \sum_{34} V_{3\tilde{4}\tilde{1}2} c_{43} n_3 s_{4} s_{1} +
\sum_{34} V_{1\tilde{2}\tilde{4}3} c_{34} s_2 s_{4} (1-n_{3})
\label{N_RPA}
\end{eqnarray}
\end{widetext}
Equation (\ref{N_RPA}) is the main N-RPA equation. It is a linear
homogenous equation for $c_{12}$. The latter has a non-zero solution
only if the determinant vanishes, which fixes the frequency
$\omega$. The normalization of $c_{12}$ is determined by
\begin{eqnarray}
1 = \langle 1_N | 1_N \rangle \approx \sum_{12} [(c^\dagger)_{21}
c_{12} n_2 (1-n_1) + (c^\dagger)_{21} c_{\tilde{2}\tilde{1}} s_1
s_2] .     \nonumber
\end{eqnarray}
Then pair-transition amplitudes between the ground state and the
excited state of neighboring even-even nuclei are calculated as
\begin{eqnarray}
\langle 0_{N-1} | K_{12} | 1_N \rangle \equiv \langle 0_{N-1} | a_2
a_1 | 1_N \rangle  \nonumber \\
\approx c_{2\tilde{1}} s_1 (1-n_2) - c_{1\tilde{2}} (1-n_1) s_2 .
\nonumber
\end{eqnarray}
In the above two equations approximations (\ref{fac_1b}) and
(\ref{fac_2b}) are used.

We emphasize that, unlike the conventional RPA or QRPA, in the
current theory $A |0_N\rangle \ne 0$. Considering the special case
of infinitesimal residual interaction (but pairing interaction has
normal strength), $|0_N\rangle$ would be almost $|\phi_N\rangle$,
and $A |\phi_N\rangle$ does not vanish for any one-body phonon
operator $A$ (\ref{phonon}).

\section{Degenerate Model}

We consider the degenerate model (degeneracy $2 \Omega$) with
``quadrupole-plus-pairing'' force:
\begin{eqnarray}
\epsilon_{12} = \delta_{12} \epsilon  ~,~~  V_{1234} = - G
\delta_{2\tilde{1}} \delta_{3\tilde{4}} - \kappa q_{14} q_{23} +
\kappa q_{13} q_{24}     \nonumber
\end{eqnarray}
in the Hamiltonian (\ref{H_f}). The occupation number
(\ref{rho_diag}) $n_1 = N/\Omega$, pair-transition amplitude
(\ref{kappa_diag}) $s_1 = \sqrt{n(1-n)}$, mean fields (\ref{f_diag})
$e_1 = \epsilon - n G$ and (\ref{delta_diag}) $g_1 = G \Omega s$ are
already calculated in Ref. \cite{Jia_pairing}. Consequently Eq.
(\ref{N_RPA}) becomes
\begin{eqnarray}
(G \Omega - \omega) (c_{12} + c_{\tilde{2}\tilde{1}}) = \kappa (q c
q)_{12} + \kappa (q c q)_{\tilde{2}\tilde{1}} ,     \nonumber
\end{eqnarray}
where we have used that operator $q$ is time-even ($q_{12} =
q_{\tilde{2}\tilde{1}}$). At infinitesimal $\kappa$, the N-RPA
frequency $\omega \approx G \Omega$, which is the correct pairing
gap.

\section{Zero-Pairing Limit}

In this section we show that the N-RPA equation (\ref{N_RPA}) goes
over to the usual RPA equation in the case of vanished pairing
interaction. In this case, $|\phi_N\rangle$ is a Slater determinant
with a sharp Fermi surface (F). The single-particle levels below F
are fully occupied and are empty above F. The pair-emission
amplitudes $s_1$ for $1 \ne F$ vanishes, and $s_F \approx
\sqrt{n_F(1-n_F)} = 0$. Introducing $r_{12} \equiv \langle 0_N |
R_{12} | 1_N \rangle = c_{12} n_2 (1 - n_1)$, Eq. (\ref{N_RPA})
becomes
\begin{eqnarray}
\omega r_{12} = e_{12} r_{12} - n_{12} \sum_{34} V_{1432} r_{34} .
     \label{RPA}
\end{eqnarray}
This is the usual RPA equation. $r_{12}$ is related to the
conventional $X$, $Y$ amplitudes (see Ref. \cite{Peter}) by $r_{mi}
= X_{mi}$, $r_{im} = Y_{mi}$, where $i < F$ is a hole level and $m >
F$ is a particle level.\\

In summary, a new version of number-conserving random phase
approximation is proposed following the procedure of deriving the
conventional (quasi-particle) random-phase approximation. The theory
is considered in two limits: the degenerate model (large pairing
case), and the vanished-pairing case. More applications/tests are
needed in the future.

In the next step anharmonicities could be included in the following
way. Using the ``factorizations'' (\ref{fac_1b}), (\ref{fac_2b}) and
their extensions, we could calculate the Hamiltonian matrix in the
collective subspace $\langle m | H | n \rangle \equiv \langle 0_N |
A^m H (A^\dagger)^n | 0_N \rangle$. Then we diagonalize the matrix
$\langle m | H | n \rangle$. Odd-mass nuclei could be treated
similarly by calculating the matrix $\langle m j | H | n j' \rangle
\equiv \langle 0_N | A^m a_{j} H a_{j'} (A^\dagger)^n | 0_N \rangle$
using the ``factorization''.

It may also be possible to generalize the conventional
particle-particle RPA (see Ref. \cite{Peter}) by treating the
Heisenberg equation of motion $[K_{12} , H]$ of the density matrix
operator $K_{12} \equiv a_2 a_1$ in a way similar to that of Eq.
(\ref{exact_eom}).

\end{document}